\documentclass{ws-procs9x6}

\begin{document}

\title{Generalized parton distributions for weakly bound systems from light-front quantum mechanics}

\author{B.~C.~Tiburzi}

\address{Department of Physics\\
University of Washington \\ 
Box $351560$ \\
Seattle, WA $98195$-$1560$\\ 
E-mail: bctiburz@u.washington.edu}


\maketitle

\abstracts{We present generalized parton distributions for weakly bound
systems on the light cone in order to build intuition about the light-front
formalism. Physics at the crossover is reviewed in terms of the light-cone Fock
space representation. Furthermore, we link light-cone Fock components to 
the equal (light-cone) time projection of the covariant Bethe-Salpeter amplitude.
Continuity of the distributions arises naturally in this weak binding model.}

\section{Introduction}
Recently much activity has been geared to investigating the next generation of exclusive processes at high momentum 
transfer\cite{Ji:1998pc}. Deeply virtual Compton scattering (DVCS) represents perhaps the cleanest of these to 
describe theoretically. The DVCS amplitude is a convolution of a hard scattering part and a set of
new structure functions, which, at leading twist, are the generalized parton distributions (GPDs). The GPDs are 
off-diagonal matrix elements of bilocal field operators which interpolate between the inclusive physics of 
parton distributions and the exclusive limit of electromagnetic form factors. Geometric and optical interpretations
of GPDs have elucidated some of the new physics they encode\cite{Burkardt:2000za}.

The purpose of this talk is two-fold. Firstly we wish to provide an intuitive
basis for the light-front formalism. Light-cone correlations appear in hard processes
and thus these correlators have a simple interpretation in terms of light-cone wavefunctions.
Secondly we present only simple models of light-cone wavefunctions in order to demonstrate
some general properties of GPDs, for example. Comparatively little work has been done
to show the light-cone Fock representation of GPDs is consistent with known symmetries
and limits. 

The organization is as follows. We begin with the simplest example of a light-cone wavefunction: 
a valence quark model for the proton. This model is shown to be insufficient for
describing beam-spin asymmetries which are sensitive to GPDs at the crossover
between kinematic regions. Higher Fock components are needed and are derived in a 
weak binding, spinless model. These GPDs are continuous at the crossover due to 
simple relations between Fock components at vanishing plus-momentum. 

\section{Valence quark models}
One can construct a relativistic valence quark model of a hadron by simply modeling 
the lowest light-cone Fock component. We analyzed GPDs from \emph{ad hoc} valence 
two-body wavefunctions earlier\cite{Tiburzi:2001ta}. For the proton, one would truncate the Fock space to 
exclude all but the three-body Fock component. To determine this constituent quark wavefunction of the proton, 
one would solve a quantum mechanical bound-state equation of the form
\begin{equation} \label{bseq}
\Bigg[ \sum_{i = 1}^{3} \frac{\mathbf{k}^\perp_{i}{}^{2} + m_{i}^2}{2 x_{i}} + \hat{V} \Bigg] 
\psi_{3}(x_{i},\mathbf{k}^\perp_{i}) = M^2 \; \psi_{3}(x_{i},\mathbf{k}^\perp_{i}),
\end{equation}
where the $m_{i}$ are constituent quark masses, and $\hat{V}$ is some effective interaction. Since 
solving a three-body equation is often quite difficult, the form of the wavefunction is usually
postulated. The functional form used is not completely arbitrary; there are minimal physical restrictions 
to be satisfied. The restriction which concerns us below is at a vanishing plus-momentum fraction: $x_{j} = 0$. 
From Eq.~(\ref{bseq}), the kinetic term as well as the interaction $\hat{V}$ force the valence
wavefunction to vanish, i.e.~$\psi_{3}(\ldots,x_{j}=0,\ldots) = 0$. 

As is known, such valence models lead to reasonable success at low energy phenomenology, particularly
hadronic spectroscopy and form factors. We shall see below that such models are insufficient for
modeling the GPDs. 

\section{At the crossover}
DVCS probes a light-cone correlation of the quark fields. Not surprisingly the GPDs can be expressed most simply
as a sum of Fock component overlaps\cite{Diehl:2000xz}. The most direct way to experimentally 
access the GPDs from the Bethe-Heitler background is through beam asymmetries, 
such as the beam-spin asymmetry\cite{Diehl:1997bu}. Additionally the beam-spin asymmetry
directly probes the Fock component overlaps since only the imaginary part of the amplitude is
needed\footnote{This is in contrast to the charge asymmetry which is sensitive to the real part of the Compton amplitude.
The real part involves an integral of the GPDs weighted by the hard scattering amplitude.}
\begin{eqnarray}
\Im \mathcal{M} &\propto& \frac{\sqrt{1-\zeta}}{1 - \frac{\zeta}{2}} H(\zeta,\zeta,t) - \frac{\zeta^2}{4 ( 1 - \frac{\zeta}{2}) \sqrt{1 - \zeta}} 
E(\zeta,\zeta,t) \\
&=& \sum_{n,\lambda_{i}} \sqrt{1 - \zeta}^{2 - n} \int_{\{n\}} \delta( \zeta - x_j) \psi_{n}^{\uparrow}{}^{*}(x^{\prime}_{i},\mathbf{k}^\prime_{i}{}^\perp,\lambda_{i}) \psi_{n}^{\uparrow}(x_{i}, \mathbf{k}^\perp_{i},\lambda_{i}) \label{sumoffock}
\end{eqnarray}
where $\int_{\{n\}} = \int \prod_{i=1}^{n} \frac{dx_{i} d\mathbf{k}^\perp_{i}}{16 \pi^3} 16 \pi^3 \delta( 1 - \sum x_{i}) \delta( \sum \mathbf{k}^\perp_{i})$
and the primed variables are given by: $x^{\prime}_{j} = 0$, $\mathbf{k}^{\prime}_{j}{}^\perp =  \mathbf{k}^\perp_{j} - \mathbf{\Delta}^\perp$ for the 
struck quark and $x^\prime_{i} = \frac{x_{i}}{1-\zeta}$, $\mathbf{k}^{\prime}{}_{i}^\perp =  \mathbf{k}^\perp_{i} + \frac{x_{i}}
{1 - \zeta}\mathbf{\Delta}^\perp$ for the spectators.
Above the final state Fock components are evaluated at a vanishing momentum fraction. Thus modeling the physics in the
lowest Fock state forces the Compton amplitude to have a vanishing imaginary part. Higher Fock components are needed for
a non-vanishing imaginary part, since they generally do not vanish for zero plus-momentum.\footnote{This point is clear if we utilize the forward limit: 
$\{\zeta, t\} \to 0$, for which the sum in Eq. (\ref{sumoffock}) reduces to the quark distribution function at zero plus-momentum. In a valence model,
the quark distribution function vanishes at this point whereas realistically the higher Fock states lead to a non-zero value (which is of course
sizeable at a high scale).} 

Not surprisingly valence quark models that match electromagnetic form factors fail to capture the physics of GPDs. As such, valence models are poor 
interpolators between the exclusive limit of form factors and the inclusive physics of quark distribution functions. A popular, simple \emph{Ansatz}
accomplishes this interpolation trivially:
\begin{equation}
H(x, \zeta = 0, t; Q^2) \sim q(x;Q^2) F(t),
\end{equation}
where $F(t)$ is the dipole fit to the form factor, and $q(x; Q^2)$ is a fit to the quark distribution function from deep inelastic scattering data\footnote{Here we include the scale dependence of the GPD which can conveniently be absorbed into the \emph{Ansatz} via the quark distribution function. In this way, 
the GPD will evolve while the form factor, which stems from a conserved current, will not.}.
This form is too simple, although it respects the properties of GPDs and is hence difficult to improve upon\footnote{This \emph{Ansatz} has been
augmented by additional physics, such as pion-pole contributions \emph{etc}., and has been used for quite complete phenomenological estimates of cross sections including 
twist-three contributions\cite{Belitsky:2001ns}.}. The full Fock space expansion, however, contains the physics of both limits and the 
light-cone wavefunctions are hence the true interpolators (from which no factorized \emph{Ans\"atze} are discernible).

\section{Weak binding model}
Although exact, the Fock space representation of DVCS is not entirely useful since non-perturbative solutions to the $N$-body wavefunctions 
are obviously not currently available. Since properties of GPDs are largely unexplored in terms of the Fock space representation, we adopt 
a two-body perturbative model to gain some insight into the problem. To deal with ambiguities (particularly the so-called non-wavefunction vertices),
we start from a covariant framework and project onto the light-cone by integrating out the light-front energy dependence. Schematically 
$\int dk^-$ projects onto the surface $x^+ = 0$. For scattering states, it has been shown\cite{Ligterink:1994tm} that projecting
covariant perturbation theory onto the light-cone results in light-cone time-ordered perturbation theory, including the delicate issue 
of renormalization.

On the other hand, for two-body bound states, a three-dimensional reduction scheme has been found\cite{Sales:1999ec} that reproduces light-front time-ordered kernels 
in the Bethe-Salpeter formalism. We recently applied this reduction scheme to bound-state matrix elements of the electromagnetic current\cite{Tiburzi:2002mn}
from which one can extract GPDs. 

The model consists of a two-body bound state (completely devoid of spin) in the ladder approximation. The bound state can be described covariantly by 
the Bethe-Salpeter wavefunction $\Psi(k,P)$, where $P$ labels the total four-momentum of the system. The two-body Fock component
is merely the projection of the Bethe-Salpeter wavefunction onto the light-cone
\begin{equation}
\psi_{2}(x, \mathbf{k}^\perp_{rel}) \sim \int dk^- \Psi (k,P),
\end{equation} 
where the relative transverse momentum is $\mathbf{k}^\perp_{rel} = \mathbf{k}^\perp - x \mathbf{P}^\perp$ and $x = k^+/P^+$. The GPD is then 
extracted from the integrand of the form factor $F$, which is a matrix element of the electromagnetic current $J^\mu$ between Bethe-Salpeter vertices. 
\begin{equation}
F \sim \int d^4 k \cdots \to \int dk^+ \delta(x P^+ - k^+) dk^- d^2 \mathbf{k}^\perp \cdots
\end{equation}
The inserted delta function above keeps the plus-momentum of the struck quark fixed; the GPD is what remains of the integral. 
Additionally performing the $k^-$ integral allows us to write expressions in terms of light-front wavefunctions in light-front 
time-ordered perturbation theory. 

\begin{figure}[th]
\centerline{\epsfxsize=3.5in \epsfbox{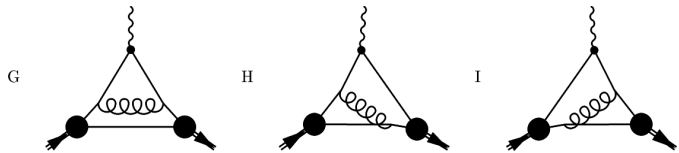}}   
\caption{Next-to-leading contributions to the GPD in the region $x > \zeta$. \label{diag}}
\end{figure}

\begin{figure}
\centerline{\epsfxsize=2.2in \epsfbox{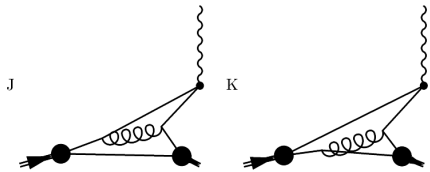}}   
\caption{Leading contributions to the GPD for $x<\zeta$. \label{ndiag}}
\end{figure}

To zeroth order in the coupling, the GPD is an overlap of two-body wavefunctions that generalizes the Drell-Yan formula
to a frame in which $\Delta^+ \neq 0$. This expression is non-vanishing only in the region $x > \zeta$. The contributions at first-order
for $x>\zeta$ are depicted in Figure \ref{diag}. Each contains a mediator that cannot be absorbed into the initial or final state vertices. 
Including the diagram in which the spectator quark has a one-loop self-interaction, we see the next-to-leading contributions have the form
of a three-to-three overlap (consistent with $x > \zeta$ contributions being diagonal in Fock space). 
Indeed the form of the three-body wavefunction is exactly as we would write down directly from time-ordered perturbation theory. 
In the region $x < \zeta$, the leading contribution is at first-order in the coupling. The relevant diagrams are shown in Figure \ref{ndiag}
and are four-to-two overlaps. These diagrams cannot be summed into an effective non-wavefunction vertex\footnote{If the four-dimensional kernel
were really three-dimensional\cite{Tiburzi:2001je}, i.e.~independent of light-front energy (instantaneous in light-cone time), 
one could appeal to crossing since true higher Fock states would be absent.} 
since the expressions do not have the same functional dependence (and hence have differing covariance properties). 
 
Continuity of the GPD at $x = \zeta$ is required by factorization\cite{Radyushkin:1997ki}. If the GPDs were discontinuous, the Compton 
amplitude would be logarithmically divergent. This weak binding model's GPD is continuous. As expected, the valence overlap vanishes at
the crossover (similarly does diagram $H$). The Born terms ($G$ and $J$) match up at $x = \zeta$ due to the dominance of the rebounding 
quark's energy in the time-ordering. For this reason, the final-state iteration diagrams ($I$ and $K$) also match at the crossover. Furthermore, 
the GPD does not vanish at the crossover, which naturally stems from higher Fock states. 
In this model, continuity at the crossover generalizes easily to all orders in time-ordered perturbation theory. 

\section{Conclusion}
Above we have seen the structure of GPDs in a simple weak binding model. Continuity of the distributions 
was maintained naturally by simple relations between Fock components evaluated at a vanishing plus-momentum 
fraction. In QCD, however, the relations between light-cone Fock components at small-$x$ are far 
richer\cite{Antonuccio:1997tw}. We have yet to see how the sum rule, which relates the zeroth moment 
of the GPD to the electromagnetic form factor, arises from relations between Fock components. This should
first be tackled perturbatively. As to constructing phenomenological models of GPDs from 
light-front wavefunctions, we have demonstrated that valence models are insufficient which suggests either that
constituent quark substructure needs to be added or that higher Fock components need to be modeled. Proceeding
on either course is not simple: continuity and polynomiality conditions are serious constraints. 

\section*{Acknowledgments}
G.~A.~Miller provided much insight throughout the course of this project. 
This work was funded by the U.~S.~DOE, grant: DE-FG$03-97$ER$41014$.

\end{document}